\documentclass{article}

\usepackage{microtype}
\usepackage{graphicx}
\usepackage{subfigure}
\usepackage{booktabs} 

\usepackage{hyperref}

\usepackage[accepted]{article}

\begin{document}

\twocolumn[
\sysmltitle{FusionStitching: Deep Fusion and Code Generation for Tensorflow Computations on GPUs}



\begin{sysmlauthorlist}
\sysmlauthor{Guoping Long}{Alibaba}
\sysmlauthor{Jun Yang}{Alibaba}
\sysmlauthor{Kai Zhu}{Alibaba}
\sysmlauthor{Wei Lin}{Alibaba}
\end{sysmlauthorlist}

\sysmlaffiliation{Alibaba}{Alibaba Inc.}

\sysmlcorrespondingauthor{Guoping Long}{guopinglong.lgp@alibaba-inc.com}

\sysmlkeywords{Code Generation, Compiler, Machine Learning, Tensorflow, GPU}

\vskip 0.3in

\begin{abstract}
In recent years, there is a surge on machine learning applications in industry.
Many of them are based on popular AI frameworks like Tensorflow, Torch, Caffe, or MxNet, 
etc,
and are
enpowered by accelerator platforms such as GPUs. One important challenge of
running Tensorflow computations on GPUs is the fine granularity problem, namely,
FLOPS of individual ops are far from enough to fully exploit the computing power
of underlying accelerators.
The XLA framework provides a solid foundation to explore this problem further.
In this paper, we propose FusionStitching, a novel, comprehensive Op fusion and
code generation system to \emph{stitch} computations into large GPU kernels. 
Experimental results on four public models and two of our large inhouse applications
show another 55\% (geometric mean) reduction of GPU kernel launches, 
compared to the XLA fusion baseline. This increases the E2E performance 
of both of our latency critical inhouse applications up to 20\%.
\end{abstract}
]



\printAffiliationsAndNotice{}  

\section{Introduction}
\label{submission}
Recent years in industry, there is a boom of machine learning applications in a diversified range of
scenarios, including images, speech/audio, NLP, CTR prediction, search and recommender systems built
on commodity graphs at billions even trillions of scale, etc. Such workloads are generally
regular in computation, and benefit a lot from modern high performance accelerators like GPUs
or TPUs. In addition, many of such models are based on popular AI frameworks like 
Tensorflow\cite{tensorflow}, Caffe\cite{caffe}, Torch\cite{torch}, CNTK\cite{cntk}, 
or MxNet\cite{mxnet}. The challenge 
is how to optimize such workloads, to achieve as much performance as possible on modern hardware.

There are roughly two categories of computations in AI workloads. One is enabled by optimized
vendor libraries, in particular MatMuls or 2D/3D convolutions and their variants. The other category 
includes foundational tensor operators, elementwise computations, memory layout transformations, and 
other workload specific ones. In order to understand the relative importance of MatMul/Convolution 
computations on various models, we collected data from 53,470 models on PAI (Platform for Artificial 
Intelligence) at Alibaba. Depending on application domains and hardware platforms, this number 
ranges from 40\% to 70\%. Therefore, computations other than Matmul/Conv deserve serious investigation 
to achieve decent performance. 
This work focuses on 
computation efficiency of this category on GPU platforms. 
\begin{figure}
\includegraphics[scale=.5, bb=75 223 536 568]{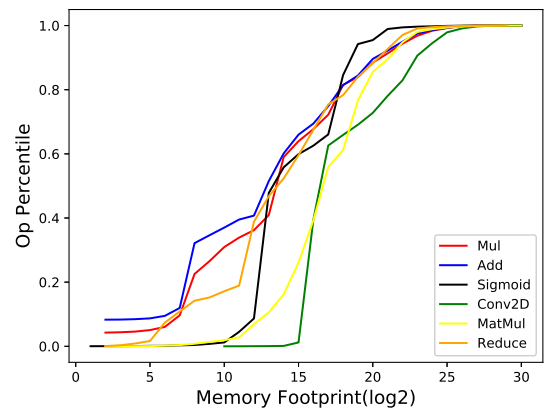}
\caption{Memory Footprint Distribution of Most Popular Ops}
\label{fig:OpHist}
\end{figure}

A well known challenge of Tensorflow computations is op granularity. Figure \ref{fig:OpHist} shows
the accumulated percentile distribution of memory footprints of six most frequent computing ops of
those 53470 models collected from PAI. Note that the \emph{reduce}(orange) line denotes the collective statistics
of four ops, namely mean, sum, min, and max, while other lines represent individual ops. The x-axis shows
the memory IO footprint size (in number of floats) measured at logarithm scale (base=2), the bigger the better. 
As can be shown, while memory footprints of MatMul/Conv2D are generally larger than elementwise/reduce ones,
most op instances have small memory footprints. Therefore, optimizations are vital to fully leverage computing 
resources and memory bandwidth.

One way to solving this problem is replacing those fine grained ops with a pre-implemented coarse grained 
version\cite{NMT_on_TVM_TF}. However, this 
approach is not scalable and incapable of adapting to diversified and fast evolving workload characteristics.
The more principaled solution is compiler driven op fusion, tuning and code generation. 
One possibility is to seperate
the schedule specification and implementation. By offloading the optimization plan specification to the user,
the runtime focuses on details of optimization realization. Prior explorations have shown promising results in
some application domains, such as Halide on image processing\cite{halide}, loopy\cite{loopy}
and TVM\cite{tvm} on array and machine
learning computations. To make this layered approach work in practice, we believe the performance insight into the
target workload domain is essential. With such insights, and given a properly defined workload model and 
target platform, it is possible to go a step futher, by allowing the user to specify computation only, rather 
than implementation schedules, as is the case in Tensorflow.

In Tensorflow, the fast evolving XLA framework provides a sound foundation to explore this problem
further. XLA's approach of partitioning a Tensorflow graph into compilable clusters, and transforming them into
concise and compact \emph{HloModules}, opens up broad possibilities to fuse, transform, and optimize computation 
kernels for GPUs.

\begin{figure}
\centering
\includegraphics[scale=.6, bb=67 50 424 209]{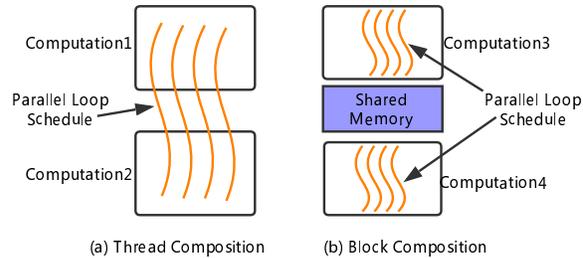}
\caption{Computation Composition}
\label{fig:composition}
\end{figure}

Current XLA op fusion algorithm, either \emph{GpuInstructionFusion} or \emph{MultiOutputFusion}, relies on a
set of static \emph{ShouldFuse} rules in order to produce supposedly profitable larger kernels. While these
rules discern fusion opportunities in many cases,
it is usually compromised by exceptions, such as expensive elementwise ops, column reductions,
batched matmuls, or memory layout transposes. In addition, at the code generation phase, XLA requires all ops 
in the fused
computation to fit into a single parallel loop emitter, and leverages the \emph{elemental\_ir\_emitter} to
compose computations of previous ops into the root loop body. In this approach, all ops in the fused 
computation in essence must share the same implementation schedule. We call this approach thread composition,
as shown in Figure~\ref{fig:composition}(a).

This work proposes \emph{FusionStitching}, a deep fusion and code generation system.
One key feature of our system is block composition at the codegen phase, as shown in 
Figure~\ref{fig:composition}(b). To support this, we propose another ir\_emitter, \emph{IrEmitterStitched}, 
to \emph{stitch} multiple computations together. In theory, we allow each computation to have its own parallel
loop emitter, and use on chip shared memory (scratchpad) as intermediary between producing and consuming 
computations. In practice, however, due to the scarsity of shared memory space, 
we leverage thread composition together with block composition. 
\begin{figure*}
\centering
\includegraphics[scale=.5]{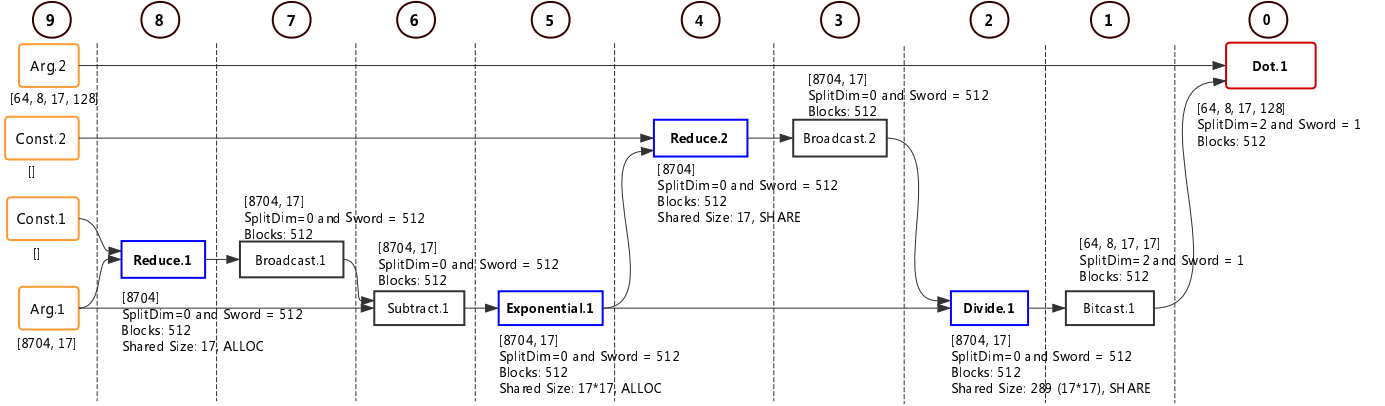}
\caption{A Motivating Example}
\label{fig:motivating_example}
\end{figure*}

This opens up further optimization tradeoffs and fusion opportunities. At the op fusion phase, due to the 
codegen capability enabled by block composition, we can further relax constraints that impose
on XLA, and thereby enabling much larger kernel granularity. Besides, to cope with much bigger implementation 
space of larger kernel sizes, a systematic approach is desirable to specify, optimize and tune kernel 
optimizations. Specifically, we make the following contributions:
\begin{itemize}
\item[-] We propose a novel deep fusion algorithm based on a layered nodes structure along the span (critical
path) of the graph. Using critical path reduction as the driving heuristic, we consider not only 
producer/consumer fusion opportunies, but also fine granularity ops that occur in the same layer, in
order to enlarge kernel granularity and reduce GPU launch overheads.
\item[-] We propose a comprehensive mechanism to specify implementation schedule space,
resolve schedule constraints, tune the search space and generate the final implementation plan,
represented as a set of launch/schedule parameters.
\item[-] We propose IrEmitterStitched, another ir\_emitter to support block level computation composition. The
core part of this ir\_emitter is a shared memory planning algorithm which orchestrates the code generation of
individual ops within the fused computation.
\end{itemize}

This paper is organized as follows. Section \ref{section:motivation} discusses motivation and overview
of \emph{FusionStitching}. Section \ref{section:fusion} presents our deep fusion algorithm. Section
\ref{section:launch} discusses schedule specification, tuning and determination of optimized implementation
plans. Section \ref{section:codegen} presents details of code generation and the shared memory management 
algorithm. Section \ref{section:evaluation} shows experimental results. Section \ref{section:relatedwork}
discusses related works and Section \ref{section:conclusion} concludes this work.
\section{Motivation and System Overview}
\label{section:motivation}
\subsection{The Motivating Example}
We consider the fusion and code generation problem of general sub-graphs that consist of four types
of ops: (1) Elementwise; (2) Shape modulation ops, such as \emph{Reshape}, \emph{Bitcast}, \emph{Transpose},
etc; (3) Reduction; (4) BatchMatMul. We include \emph{BatchMatMul} because in some of our critical
production models, such ops usually involve workload specific shapes, and cuBLAS kernels do not 
deliver satisfactory performance.

Figure \ref{fig:motivating_example} shows a motivating example. We arrange ops in a layered structure 
(denoted as circled numbers), with 
\emph{layer $9$} being input ops (top), and \emph{layer $0$} being output 
(BatchMatMul (\emph{Dot.1}) in this case). In 
complex graphs, such a layered structure proves to be very useful to fusion decision making. Black arrows
(from left to right) show data dependances. Next to each op, there
are annotating texts that show important schedule/code generation attributes. The (\emph{SplitDim}, 
\emph{Sword}) pairs are schedule parameters used for tuning the implementation space (Section 
\ref{section:fusion}). One use of this
pair of parameters is to decide \emph{Blocks}, the number of thread blocks (CTA) used for computing this op. 
The \emph{Shared} attributes associated with \emph{reduce}, \emph{exponential}, and \emph{divide} are 
related to on chip shared memory management. Together with the \emph{Size} attribute, \emph{ALLOC} or
\emph{SHARE} mean we need to allocate space for the current op, or reuse a buffer allocated for a previous
op, respectively.

Our system provides the capability to fuse and generate optimized code for the entire graph.
Whether it's beneficial to fuse \emph{dot} depends on workloads. In some
of our inhouse workloads, the batched dot shape is too marginal to get any benefit calling cuBLAS, and
this pattern happens to be the core part of an inner loop body. In this case, fusing everything proves
to be very useful. In general, we leave the decision of whether to fuse \emph{BatchMatMul} to the user.
\subsection{FusionStitching: The System Overview}
\begin{figure*}
\centering
\includegraphics[scale=.6]{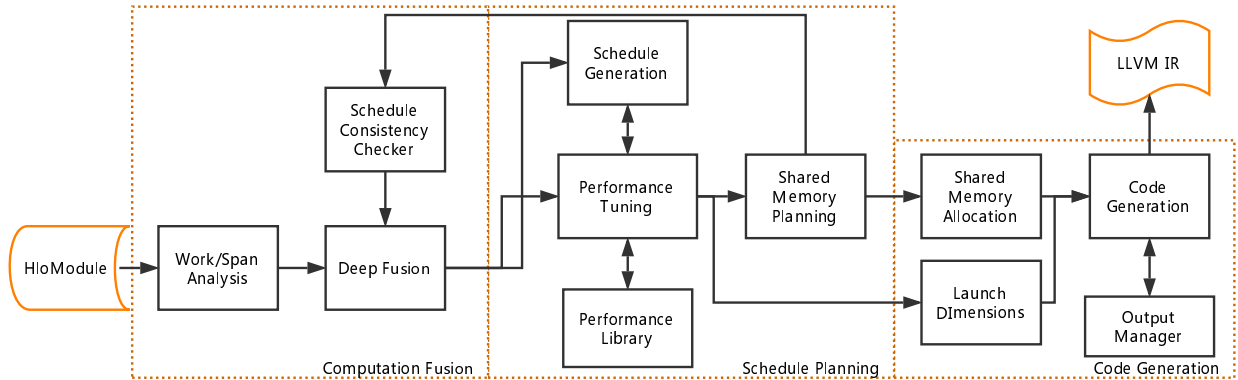}
\caption{The System Overview}
\label{fig:overview}
\end{figure*}

The system overview is shown in Figure \ref{fig:overview}. On the high level, the system takes a \emph{HloModule}
as the input, passes three stages of processing (op fusion, schedule planning and code generation), and finally 
generates the LLVM IR.

In the computation fusion stage, we first perform a Work/Span (critical path) analysis\cite{workspan}, and allocate 
a layer number (as shown in Figure \ref{fig:motivating_example} for each op according to its depth in the span. 
Then starting from the \emph{root} (such as \emph{Dot.1} in Figure~\ref{fig:motivating_example}), we fuse ops 
iteratively across different span layers, as long as
the fusion decision passes the schedule consistency check. The fusion process iterates until no fusion opportunity
is available. Then the transformed \emph{HloModule} is passed on to schedule planning.

The implementation space for a large fused computation can be huge. 
Schedule planning searches a domain driven, well defined schedule space for optimized implementations.
It takes a fusion plan as input, and generates optimized schedule parameters, shared
memory usage plans and launch dimensions for the following code generation phase. It also provides
performance heuristics regarding current fusion plan as feedback information to \emph{ScheduleConsistencyChecker}.
The scheduling process involves four important submodules: schedule generation, performance library, tuning and
shared memory planning, which will be presented in detail in section \ref{section:launch} and 
section \ref{section:codegen}.

Based on schedule parameters, the code generation pass finally allocates shared memory, set the kernel
launch dimension, and emits LLVM IR code for each op according to its tuned parameters. Note that the use of
shared memory as an intermediary among ops is important to achieve decent performance. One example is shown
in Figure \ref{fig:motivating_example}. If no shared memory, expensive ops like \emph{exponential} and 
\emph{divide} can only be composed through thread composition. In this case, computation of these ops will 
be nested into the inner
loop of the root \emph{dot} op, causing notable performance degredation due to the duplicated computation. 
With shared memory, threads within a thread block for different ops can cooperate differently, thus opens up more optimization opportunities
(section \ref{section:codegen}).
\section{Deep Fusion}
\label{section:fusion}
\subsection{The Work/Span Analysis}
Work/Span analysis is a handy tool to analyze parallel work and the critical path of computation 
graphs\cite{workspan}. In our implementation, we assign a number, the \emph{span} to each instruction of the
\emph{HloModule}. First, the \emph{root} instruction have zero span. For any other instruction, its span
equals the maximum span of its users plus one. 
Work/Span analysis is also useful to profile entire Tensorflow graphs. However, standard Work/Span analysis
works fine only when the graph is absent of dependancy loops. It is not uncommon for practical Tensorflow
graphs to include large, possibly nested while loops. In this case, we perform an preprocessing
step to partition all nodes into multiple subgraphs, each belonging to a separate frame context. 
We then perform Work/Span analysis for each frame context independently.

After running this analysis, the maximum \emph{span} assigned is the length of the critical path.
Instructions with the same \emph{span} are on the same layer 
(as shown in Figure~\ref{fig:motivating_example}),
and there are no data dependances among them. Next we present our fusion algorithm that leverages this
information to enlarge the kernel granularity by effectively reducing the \emph{span} of the computation.
\subsection{The Fusion Algorithm}
Through Work/Span analysis, we partition all instructions within a module into numerous layers (as shown in
Figure~\ref{fig:motivating_example}), where instructions in each layer have the same span. 
Today most AI models rely on library calls to perform
MatMul/Conv. Since we do not fuse across library calls, we are interested in computation 
subgraphs that exist in between any two consecutive library call layers (LC-layer). The basic intuition 
of our fusion algorithm is to fuse as many instructions as possible, considering various fusion constraints,
in the subgraph of computations between two LC-layers.

Starting from a given LC-layer, up to the next LC-layer (\emph{roof}), for each layer (denoted as the
\emph{root} layer in discussions below), repeat the following two steps. First, we perform an intra
layer \emph{ElementwiseFusion} at the root layer, resulting in a set of fused computations. Second, for each
fusion instruction (\emph{fusion\_root}) in the root layer, we use Algorithm~\ref{alg:deepfusion} to 
perform sub-graph fusion up to the next \emph{roof}.

\emph{ElementwiseFusion} targets intra layer nodes without producer/consumer relationships.
The primary target is for small weight accumulation layers which occur 
frequently in training graphs. For a large number of such fine grained (eg. $< 10us$) kernels, 
fusing them together can reduce substantive launch overheads. The exact number of fused computations to generate 
depends on two factors.
One is schedule compatiblity. In practice, elementwise instructions within a layer naturally fall into
a few groups according to output shapes. We will discuss more on schedule planning in the next section.
The other factor is the fused memory footprint. We use a tunable threshold parameter to control the fusion
granularity, in order to avoid extra large elementwise computations with too many outputs. 

One implementation of subgraph fusion, starting from \emph{fusion\_root}, up to the next \emph{roof}, 
is shown in
Algorithm~\ref{alg:deepfusion}. The map \emph{hlo\_span} is the result of the Work/Span analysis, recording
the span for all instructions. The algorithm traverses instructions layerwise, starting from the next layer,
up to the \emph{roof}. During this traversal, instructions are fused (put in the \emph{fused} set), 
or gave up (put in the \emph{giveup} set). 

The procedure \emph{SchdConsistent} decides whether or not to fuse an instruction \emph{hlo} with the
\emph{fusion\_root}. First, it checks if \emph{hlo} has a user in the \emph{giveup} set. If so, fusion
stops in order to avoid potential cyclic dependance loops. Second, it checks if \emph{hlo} has a user in
the \emph{fused} set. If not so, fusion stops because we consider producer/consumer fusion only here, and 
leave the other case to \emph{ElemwiseFusion}, as discussed above. Finally, it checks if it's possible
to resolve an optimized schedule for the fused computation, and stops fusion if not. We will discuss 
schedule planning and optimization in more detail in Section~\ref{section:launch}.
\begin{algorithm}[tb]
   \caption{Subgraph Fusion Algorithm}
   \label{alg:deepfusion}
\begin{algorithmic}
   \STATE {\bfseries Input:} $fusion\_root$, $roof$, $hlo\_span$
   \STATE Initialize $curr\_span = hlo\_span[fusion\_root]$
   \STATE Initialize empty set $fused$
   \STATE Initialize empty set $giveup$
   \FOR{$l=curr\_span+1$ {\bfseries to} $roof-1$}
   \STATE Initialize $hlos$ be all instructions with $span == l$
   \FOR{$hlo : hlos$}
   \IF{$SchdConsistent(fusion\_root, hlo, fused, giveup)$}
   \STATE $Fuse(fusion\_root, hlo)$
   \STATE $fused.insert(hlo)$
   \ELSE
   \STATE $giveup.insert(hlo)$
   \ENDIF
   \ENDFOR 
   \ENDFOR 
\end{algorithmic}
\end{algorithm}
\section{Schedule Planning}
\label{section:launch}
\subsection{Schedule Specification}
In a fused computation, each instruction has an output shape, which defines the total work space.
The implementation space for each instruction can be huge. It is prohibitively
expensive to exhaust the entire space in order to get the most optimized kernel. Before we discuss 
our tradeoff on this issue, let's reiterate two major objectives of op fusion: (1) reduce the memory
footprint of the fused computation; (2) reduce the number of kernel launches due to the fine grained
nature of many Tensorflow ops. In practice, most fused computations in our workloads are memory
intensive, elementwise computations. Thus our design rationale is to facilitate the composition
of numerous instructions within a kernel in order to fully take advantage of hardware resources, 
rather than pursuing extreme performance of individual ops.

For each instruction, we define three parameters on the output shape (the work space) to fully 
specify an implementation schedule: $split\_dim$, $sword$, and $sched\_type$. The idea is to split
the work space into multiple data chunks, where each thread block (CTA) works on a chunk. Here, 
$split\_dim$ denotes a dimension where we split the work space. $sword$ denotes how we partition 
the dimension $split\_dim$. $sched\_type$ can be either \emph{Row} or \emph{Column}.
\begin{figure}
\centering
\includegraphics[scale=.39, bb=58 47 645 298]{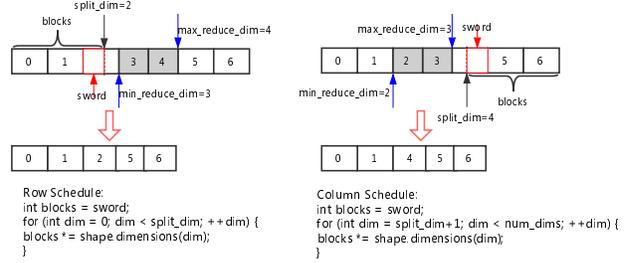}
\caption{An Example of Schedule Space for \emph{Reduce}}
\label{fig:schedule}
\end{figure}

As an example, Figure~\ref{fig:schedule} shows a \emph{Row} schedule (left) and  a \emph{Column}
schedule (right) for \emph{Reduce}, both reducing input tensors of $7$ dims to output tensors of
$5$ dims(for the left \emph{Row} schedule dimensions $3$, $4$ are reduced, while for the right 
\emph{Column} schedule dimensions $2$, $3$ are reduced, which correspond to the gray cells in the figure). 
$blocks$ denotes the number of GPU thread blocks, or data chunks after partitioning the
entire work space. In \emph{Row} schedule, we use dims on the left (more significant) of 
$split\_dim$ as $blocks$. The C code excerpt shows details on how to calculate $blocks$ using 
$split\_dim$,$sword$, and shape dims.

Let $num\_dims$ be the number of dims for a tensor shape. $split\_dim$ must be integers in range 
of $[0, num\_dims)$. Let $K$ be the the size of dimension $split\_dim$.
$sword$ must be a divisor of $K$.
$sched\_type$ can be either $Row$ or $Column$. Given a hlo instruction, the Cartesion product
of legal value sets of $split\_dim$, $sword$, and $sched\_type$ defines the entire schedule space.

The size of the schedule space of a single op depends on its output shape dims, but is
usually small in practice. This is important for compilation speed.
Our schedule specification has relatively small search space. It is not designed for exhausting 
all implementation possibilities.
Yet, together with tuning, it enables most important kernel optimizations we want for GPUs,
while at the same time allows simple and concise code generation implementations,
thanks to the computation regularity of Tensorflow ops.
\subsection{Schedule Constraints and Propagation}
The output shape of the fused computation is the same as that of its root (output) instruction. 
Given a valid schedule of the root instruction, we must decide whether it is satisfiable by all
other instructions of the computation. We use an algorithm conceptually similar to 
Work/Span analysis to resolve schedule constraints for all other instructions.
Note that for each instruction, the valid schedule is defined on its output shape.
If schedule constraints are satisfiable, we re-iterate and back propagate the schedule to its
input shape(s), which correspond to the output shape(s) of its operand(s). Otherwise, the
schedule initiated from the root instruction is not satisfiable for the computation.

Figure~\ref{fig:schedule} shows an example of schedule constraints. For $Reduce$, we require all
reduction dims be in the same thread block in order to balance between codegen simplicity and
kernel efficiency. In this case, if $split\_dim < min\_reduce\_dim$, only $sched\_type = Row$
is meaningful. The case for $Transpose$ is similar. Other constraints include the divisability
requirement needed by $sword$.
\begin{table}[t]
\caption{Schedule Constraints Propagation Rules}
\label{tbl:constraint_prop}
\begin{center}
\begin{small}
\begin{tabular}{ll}
\toprule
Op & Rules \\
\midrule
Elementwise & Pass $Row$, $Column$ \\
Transpose & $split\_dim <= min\_trans\_dim$ Pass $Row$\\
          & $split\_dim >= max\_trans\_dim$ Pass $Column$\\
Reduce & $split\_dim <= min\_reduce\_dim$ Pass $Row$\\
       & $split\_dim >= max\_reduce\_dim$ Pass $Column$\\
BatchDot & $split\_dim < num\_dims - 2$ Pass $Row$\\
Reshape & Reshape transform $split\_dim$ and $sword$\\
        & Pass $Row$, $Column$\\
Broadcast & Broadcast transform $split\_dim$ and $sword$\\
        & Pass $Row$, $Column$\\
\bottomrule
\end{tabular}
\end{small}
\end{center}
\vskip -0.1in
\end{table}

Table~\ref{tbl:constraint_prop} summarizes rules of different ops for schedule constraints 
propagation.
For {Elementwise}, propagate the schedule either $Row$ or $Column$, back to its operand(s).
For \emph{Reduce}, only Propagate $Row$ schedule if $split\_dim < min\_reduce\_dim$. If
$split\_dim >= min\_reduce\_dim$, adjust $split\_dim$ and $sword$ according to reduce dims.
Propagate $Row$ or $Column$ schedule if $split\_dim = min\_reduce\_dim$. Only propagate $Column$
schedule if $split\_dim > min\_reduce\_dim$. The case for \emph{Transpose} is similar.
For \emph{BatchDot}, only $Row$ schedules are propagated and $split\_dim$ must be a 
batch dim ($<num\_dims-2$), otherwise the schedule is not satisfiable.

The \emph{Reshape} or \emph{Broadcast} modulates shapes. Therefore, we first transform the output
$split\_dim$ and $sword$ to the input $split\_dim$ and $sword$, according to schedule 
specification shown in Figure~\ref{fig:schedule}. Then we propagate the schedule, either 
$Row$ or $Column$.
\subsection{Schedule Tuning}
There is always a valid $Row$ schedule for any fused computation, with $split\_dim=0$ and 
$sword=1$. In this case, we only use one thread block for all instructions.
However, in practice this will always lead to under unitilization of GPU resources. Together
with the performance library (discussed blow), schedule
tuning iterates over all candidate schedules of the \emph{root} to look for the most
efficient one. We use this optimized schedule to direct code generation.

If the fused computation has one single \emph{root}, we iterate over all its candidate schedules.
For each schedule, we test if it is satisfiable. If true, we lookup the performance library w.r.t.
the schedule, and sum up the kernel execution time of all ops of the computation.
The schedule with the best performance is chosen for code generation.

If there are multiple \emph{roots}, we use a two-stage approach to speedup exploration
of the search space. In the first stage, we iterate over all \emph{roots}. For each \emph{root},
we compute two sets, one is the valid $blocks$ set (shown in Figure~\ref{fig:schedule}), the other
is the set of valid schedules corresponding to valid $blocks$. Once $blocks$ sets for all 
\emph{roots} are available, we perform intersection on these sets to resolve all candidates
blocks that satisfiable by all \emph{roots}. This reduces the performance tuning space that 
needs to explore next.

The second stage starts from the resultant $blocks$ set agreed by all \emph{roots}. We iterate
over all schedules corresponding to the $blocks$ set. For each schedule, we accumulate the
kernel execution for each \emph{root}, and sum them up to obtain the whole performance metric
for the computation. The schedule with the best performance is our chosen target.

In implementation we perform two additional optimizations. First, in evaluating performance
of individual instructions, we sometimes ignore those computationally trivial ops, such 
as \emph{Reshape},
\emph{broadcast}, small \emph{Transpose} ops, etc. Such ops can be inlined via thread
composition (similar to ElementalIREmitter in XLA) with negligible performance loss. Yet if
we keep them, their strict modulation of shapes sometimes rejects highly optimized schedules.
Bypassing them can make optimized schedules be satisfiable.

The second optimization is further pruning of the search space if there are multiple 
\emph{roots}. During the second stage of schedule evaluation, we always keep the best
performing schedule achieved so far. If, during the evaluation process of some schedule,
the execution time accumulated has already exceeded that of the total latency of the 
best schedule, we simply skip the process and continue to explore the next schedule.
\subsection{The Performance Library}
The performance library is a key-value store, which contains kernel performance data of 
various types of instructions under different implementation schedules. Common features included 
in a key include \emph{opcode}, \emph{shape}, \emph{split\_dim}, \emph{sword}, 
\emph{sched\_type} and \emph{thread block size}. The thread block size is an integer in 
$[1,1024]$, and must be a multiple of GPU warp size ($32$). There are also op specific
features. For instance, \emph{Reduce} (or \emph{Transpose}) has an additional feature,
\emph{reduce\_warps} (or \emph{trans\_warps}), meaning how many GPU warps in the thread block
are used to perform the reduction (or transpose) loop.

We keep the performance library in permenant storage for repeated usages. At system
initialization, the library is loaded into memory. During the tuning process, the library 
module takes schedule keys as lookup requests. If the key exists in the library, the result
is returned immediately. Otherwise, the module constructs a CUDA C kernel from the key,
compiles and executes it on the GPU. We use the \emph{nvprof} tool to collect the kernel
execution time and insert the new key-value pair to the library for future use.

When a key misses the library, the kernel generation and performance collection
seems to be costly operation during JIT compilation. This is true in the initial warmup
phase. Later on we observe high degree of data reuse in our workloads. In addition, as
discussed before, most kernels only take several to tens of microseconds to execute.
Nevertheless, it should be possible to build a learning model to predict a performance
metric from features in the key, and return the predicted value to the tuning process
immediately, thus shortening the critical path by offloading the kernel generation, 
compilation and execution asynchronously. We will leave this as future work.

The goal of fusion is to pack computations of all instructions into a single kernel. In
schedule tuning, we are using accumulated performance of individual ops to measure the
performance of the kenel of the entire fused computation. This approach does not mean to
predict exact execution time of the fused kernel, but works well in reaching an
optimized set of parameters to effectively direct code generation.

Based on the concise specification of schedule space, effectively schedule space exploration 
and performance library driven tuning mechanism, \emph{FusionStitching} can efficiently 
enumerate huge number of fusion possibilities, thus open much more opportunity for the subsequent 
code generation phase, as illustrated in Section \ref{section:evaluation} 

\section{Code Generation}
\label{section:codegen}
\subsection{Shared Memory Planning}
The on chip shared memory is essential to facilitate thread block composition of
numerous compute expensive ops. This is important to achieve relatively large and optimized
kernels. To perform shared memory planning, we first identify candidate ops which may
need shared memory, then prioritize shared memory usage to most critical ops when space
is not enough, and facilitate space sharing among ops on the data flow.
\subsubsection{Size Requirements Analysis}
Size requirements analysis identifies all ops that may use shared memory. In the example
computation shown in Figure~\ref{fig:motivating_example}, ops in green boxes have shared 
memory requirements. There are several cases to note.

One is direct allocation. For \emph{Reduce} or \emph{BatchDot}, if it is not the root 
instruction, we must allocate shared memory for its itermediate results, allowing consumer
ops to use seperate parallel loop emitters to generate code.

Other cases are related to expensive elementwise ops, such as \emph{Exp}, \emph{Divide},
\emph{Log}, etc. In general, if such an instruction has multiple users, we may want to
allocate shared memory to buffer its results in order to achieve as much computation
reuse as possible. Note that this is true even for inexpensive ops as well.
This is performance consideration. However, if size requirements have reached
a limit, we shall give up shared memory usage in a proper order in these cases, 
by recomputing those elementwise ops to ensure correctness.

For an expensive elementwise op, sometimes even if it has only one user, we must use
shared memory in order to achieve acceptable performance. One example is shown in 
Figure~\ref{fig:motivating_example}. The \emph{Divide.1} is followed by \emph{Bitcast.1},
which is then followed by a \emph{BatchMatMul} (\emph{Dot.1}). 
Due to high degree of data reuse in \emph{Dot.1},
shared memory is important for performance here. To address this issue,
we analyze data flow in this case in order to identify all such expensive ops.
\subsubsection{Size Shrinking}
Size shrinking is a technique when size requirement of the fused computation exceeds
the shared memory limit. One main reason this happens is when $blocks$ is small, where
each thread block needs to process a large chunk of data. The basic idea to this problem
is to trade shared space for recomputation. To reduce size requirements,
we start from inexpensive elementwise ops with multiple users, then expensive elementwise 
ops with multiple uses, finally expensive ops with transitive uses by \emph{BatchMatMul}.
Even if we follow this order, there may still be multiple candidates
to choose. In this case, we prioritize the one that is closest to the \emph{root} 
instruction in the span of the graph.

Size shrinking is a best effort approach to reduce shared memory usage. If, after
shrinking and space sharing analysis (discuss blow), there is still not enough space,
a feedback signal is generated back to \emph{ScheduleConsistencyChecker} in the fusion 
module to trigger other fusion decisions. In practice, this happens only on large fused
computations where schedule planning fails to produced an optimized one. 
Thus this feedback provides an effective mechanism to control fusion granularity.
\subsubsection{Space Sharing}
Space sharing is an effective technique to reuse shared memory space. As shown in
the example in Figure~\ref{fig:motivating_example}. \emph{Reduce.2} reuses shared
space allocated for \emph{Reduce.1}. \emph{Divide.1} reuses shared space allocated
for \emph{Exponential.1}.

To facilite sharing, we first build a dominance tree\cite{dominance} starting 
from \emph{root} instruction. Then we perform another round of data flow analysis 
using the dominance tree to realize space sharing. As to the case shown in 
Figure~\ref{fig:motivating_example}, shared space allocated for \emph{Reduce.1} can be 
shared after \emph{Expontial.1}, and can be reused by \emph{Reduce.2} because \emph{Reduce.2}
dominates \emph{Reduce.1}. Similarly, \emph{Divide.1} dominates and reuses the
buffer allocated for \emph{Exponential.1}.
\subsection{Code Generation}
\begin{algorithm}[tb]
   \caption{IrEmitterStitched}
   \label{alg:codegen}
\begin{algorithmic}
   \STATE {\bfseries Input:} $hlo$, $root$, $shared$, $schedule$, $generators$
   \IF{$!root$ \&\& $!shared.count(hlo) \&\& !dot \&\& !reduce$}
   \STATE return $ElementalIrEmitter(hlo)$
   \ENDIF
   \STATE $StitchedEmitter(hlo, schedule)$
   \IF{$shared.count(hlo)$}
   \STATE $EmitWriteSharedArray$
   \ENDIF
   \IF{$root$}
   \STATE $EmitWriteOutputArray$
   \ELSE
   \STATE $EmitGenerator(generators, hlo)$
   \ENDIF 
\end{algorithmic}
\end{algorithm}
The schedule and shared memory planning setup the foundation for codegen.
We build our work based on the \emph{hlo} visitor framework available in XLA.
The \emph{GpuElementalIrEmitter} in XLA implements thread composition of computations. 
Algorithm~\ref{alg:codegen} sketches the basic idea of our block composition 
procedure, \emph{IrEmitterStitched}.

There are several inputs to \emph{IrEmitterStitched}. $hlo$ is the target instruction to
emit code for. $schedule$ and $shared$ are outputs of schedule and shared memory planning,
respectively. $root$ tells if $hlo$ is the root instruction. $generators$ is similar to
the $generators\_$ map in XLA, the difference is on shared memory handling. If $hlo$ is not
the output instruction, is neither \emph{BatchMatMul} nor \emph{Reduce}, and does not use 
shared memory as well, we fallback to \emph{ElementalIrEmitter} in XLA; otherwise
\emph{StitchedEmitter} is called to emit code based on an optimized $schedule$. When this 
is done, we store computation results to shared memory if required by calling 
\emph{EmitWriteSharedArray}. If $root$ is true, which means $hlo$ is an output of the
computation, code is emitted to write results to global memory via \emph{EmitWriteOutputArray}.
If $root$ is false, we insert an entry to $generators$ map for $hlo$, in order to support
further composition of $hlo$ with other instructions. In implementation, we encapsulate
codegen logic related to computation results, including shared/global memory handling into
an \emph{OutputManager} object, as shown in Figure~\ref{fig:overview}. 
\section{Experimental Evaluation}
\label{section:evaluation}
\subsection{Experimental Setup}
We implemented \emph{FusionStitching} on Tensorflow 1.7. Experimental results are collected
on a Pascal GPU, with 3584 cores and 64KB shared memory per SM.
Table~\ref{tbl:workloads} summarizes our benchmarks, ranging from small to medium public
models to large inhouse applications in our production environments. \emph{LR}, \emph{W2V}, 
\emph{RNN} and \emph{BiRNN} are from public\cite{workloads}. all with default configurations. 
\emph{Speech} is an inhouse speech application, training voice samples collected from millions
of consumer side portable audio systems.

\emph{NMT} is an inhouse variant of neural machine translation based on the attention
mechanism\cite{attention, NMT}. There are two use cases. One
is offline translation of descriptions of billions of commodities from one language to
another. In this case, batch processing is available to maximize efficiency. The other use
case is for realtime, online communication between sellers and buyers. In this case, batch
size is small, and latency is critical. In both cases, every millisecond of performance
imporvement is of significance in practice. There is strong incentive to optimize
as much as possible beyond MatMul/Conv.

Our evaluation baseline is the XLA implementation of fusion and code generation. It is important
to note that XLA has already done excellent work on common elementwise and producer/consumer
patterns. With \emph{FusionStitching}, we are interested in how much
\emph{additional imporvement} is possible for these workloads.
\begin{table}[t]
\caption{Benchmarks}
\label{tbl:workloads}
\begin{center}
\begin{small}
\begin{tabular}{lll}
\toprule
Name & Category & Description\\
\midrule
LR & Training & Logistic Regression\\
W2V & Training & Word2Vector\\
RNN & Training & Recurrent Neural Network\\
BiRNN & Training & Bidirectional RNN\\
Speech & Training & Speech Recognition\\
NMT & Inference & Neural Machine Translation\\
\bottomrule
\end{tabular}
\end{small}
\end{center}
\vskip -0.1in
\end{table}
\subsection{Fusion Potential Analysis}
\begin{figure}
\centering
\includegraphics[scale=.5]{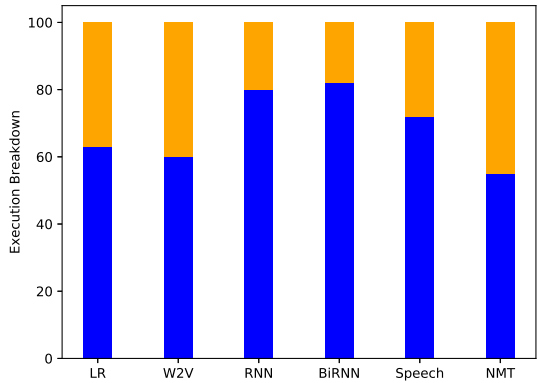}
\caption{Execution Breakdown}
\label{fig:breakdown}
\end{figure}

Optimization targets of \emph{FusionStitching} are subgraphs of ops except library calls, which
in our case only cuDNN and cuBLAS are relevant.
Figure~\ref{fig:breakdown} shows execution breakdown between \emph{MatMul/Conv} and other ops
for all benchmarks. As can be seen, the potentially fusable component (the top portion)
takes 20\% to 50\%. Large, dense \emph{MatMul/Conv} ops are friendly to GPUs, but are
computationally costly. In practice we tend to use less expensive ops for acceptable accuracy.
In addition, some \emph{MatMul/Conv} ops have particular sizes where performance gain is
very marginal to call vendor libraries. Deep fusion and efficient code generation is
critical for performance in these cases.
\subsection{The Fusion Ratio}
\begin{figure}
\centering
\includegraphics[scale=.5]{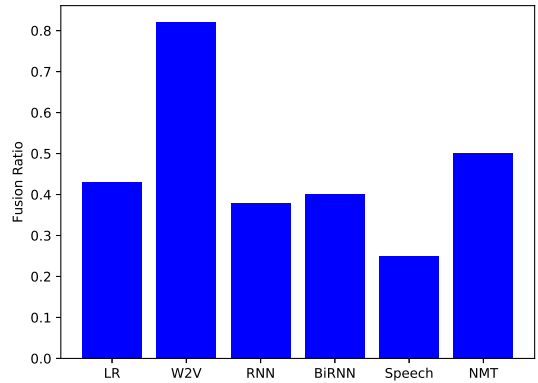}
\caption{Fusion Ratio}
\label{fig:fusionratio}
\end{figure}

One important goal of fusion is to enlarge granularity, thus reduce the number of GPU
kernels launched. We measure the ratio between the number of kernels of \emph{FusionStitching}
and that of the baseline (excluding library call kernels). We use \emph{nvprof} to collect
details of kernels information. The result is shown in Figure~\ref{fig:fusionratio}.

The fusion results depend on workloads.
For most of them, the fusion rate is less than $0.5$. This means \emph{FusionStitching}
can reduce the number kernels further to less than half the number of the baseline. \emph{W2V}
has the highest fusion ratio ($0.82$), because the core computation pattern in this case is
friendly to XLA, with limited room left for futher fusion. \emph{FusionStitching} performs
best on \emph{Speech} ($0.25$). In this case, there are complex
interaction patterns among \emph{reduce}, \emph{transpose}, \emph{concat}, and 
\emph{elementwise} ops. \emph{FusionStitching} handles them gracefully. 
\subsection{Performance Speedup}
\begin{figure}
\centering
\includegraphics[scale=.5]{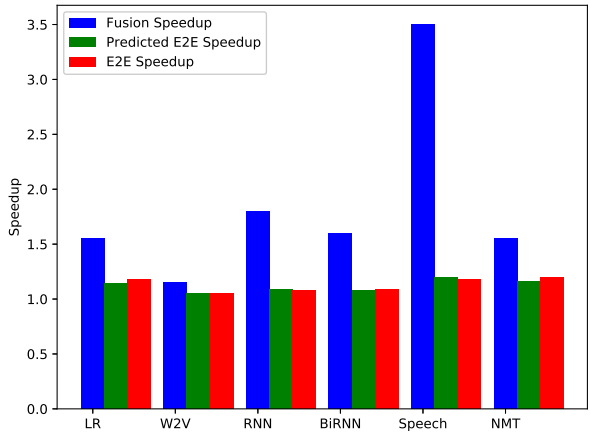}
\caption{Performance Speedup}
\label{fig:perf}
\end{figure}

The ultimate goal of fusion and code generation is for performance. Figure~\ref{fig:perf}
shows results for all workloads. We report three numbers for each benchmark. The 
$FusionSpeedup$ (left) measures performance imporvement of the fusable portion only
(in contrast to MatMul/Conv portion, as shown on the top of Figure~\ref{fig:breakdown}).
e use $FusableRatio$ to denote the execution time ratio of the fusable portion.
The end to end (E2E) speedup (right) measures performance speedup of the whole network.
The predicted E2E (middle) predicts the actual E2E speedup using the following formula:

\begin{center}$1 + FusableRatio * (1 - \frac{1}{FusionSpeedup})$\end{center}

The $FusionSpeedup$ ranges from $1.15$ (\emph{W2V}) to $3.5$ (\emph{Speech}). The average
speedup (geometric mean) is $1.74$. This speedup roughly corresponds to the reciprocal of 
the fusion ratio. The reason for this is that, in most fusion cases in these workloads,
ops are generally fine grained, memory intensive. Fusing them together effectively reduces
launch overheads and memory footprints.
This motivates us to introduce the above empirical formula to predict E2E speedup.
As can be seen, predicted speedups are close to measured E2E speedup numbers.

Besides $FusionSpeedup$ (capability measurement), $FusableRatio$ (potential measurement)
also has strong impact on E2E performance speedups. E2E speedup from \emph{FusionStitching}
varies depending on workloads, ranging from $5\%$ to $20\%$, with geometric mean $13\%$.
\subsection{Shared Memory Analysis}
In \emph{FusionStitching}, on chip shared memory is essential to composing numerous ops 
with different parallel loop emitters together.
Table~\ref{tbl:sharedmem} summarizes shared memory usage behaviors. The \emph{Average} 
column shows on average how much shared memory (in bytes) has been allcoated for each 
kernel. The \emph{Max} column shows the maximum space (in bytes) allocated. We set a
upper limit (currently 20KB) for shared memory usage of a kernel. Once the requested
size exceeds this limit, the shrinking process is triggered. The \emph{\#Shrink} column
shows how many kernels have triggered the shrinking process. Finally, the last column
shows on average, the percentage of space that is shared by multiple ops of the total
allocated space for the kernel.

Different workloads exhibit very different shared memory behaviors. \emph{LR}, 
\emph{W2V}, \emph{RNN} and \emph{BiRNN} have relatively simple producer/consumer patterns,
and neither size shrinking nor sharing happens. \emph{Speech} has large requirements
for shared memory. This is in part due to large computation granularity.
In addition, shape modulation ops (such as \emph{transpose},
etc) sometimes result in large thread block sizes in schedule planning, increasing
shared memory requirements. While little ($1\%$) allocated space is shared in \emph{Speech},
this number is $17\%$ for \emph{NMT}, indicating certain degree of computation results
reuse in the graph, as illustrated in Figure~\ref{fig:motivating_example}. The pattern in
this figure is one of the computationally intensive subgraphs of \emph{NMT}.
\begin{table}[t]
\caption{Shared Memory Statistics}
\label{tbl:sharedmem}
\begin{center}
\begin{small}
\begin{tabular}{lllll}
\toprule
Workload & Average & Max & \#Shrink & Shared Ratio \\
\midrule
LR & $64$ & $128$ & $0$ & $0.$ \\
W2V & $96$ & $288$ & $0$ & $0.$ \\
RNN & $1344$ & $3072$ & $0$ & $0.$ \\
BiRNN & $1376$ & $5120$ & $0$ & $0.$\\
Speech & $9504$ & $16416$ & $3$ & $0.01$\\
NMT & $256$ & $10432$ & $0$ & $0.17$\\
\bottomrule
\end{tabular}
\end{small}
\end{center}
\vskip -0.1in
\end{table}
\section{Related Work}
\label{section:relatedwork}
GPU kernel fusion, inspired from classical loop optimizations\cite{loopfusion, Kennedy}, 
is known to boost performance in other application domains. 
In database domain, \emph{KernelWeaver}\cite{kernelweaver} proposed 
transformations to fuse execution of multiple operators into a single kernel. This work
provided support for both thread and block (CTA) composition of operators, yet with 
little support for tuning of implementation schedules. In the HPC domain, \cite{hpcfusion}
formulated GPU kernel fusion as an combinatorial search problem, and searched the 
solution space for an optimized fused kernel. Our work targets Tensorflow computation
graphs, and proposes dedicated fusion, tuning and code generation to achieve high performance.

The parametric representation of the implementation schedule is inspired from Halide\cite{halide}
and TVM\cite{tvm, tvmgit}. However, instead of relying on users to specify schedule details, we 
propose a compact and efficient schedule specification, and tuning framework for Tensorflow 
graphs. Experimental results show decent performance gain on the fusable portion of the graph.
The layered span graph, used in our fusion algorithm, is inspired from \emph{Work/Span}
\cite{workspan} analysis of parallel computation DAGs and layered dependance graph 
representation\cite{lag} of stencil kernels.

In our work, we do not fuse dense DNN layers, and leverage vendor libraries for performance.
However, there are recent advances on code generation of fast DNN kernels.
\cite{pbqp} proposed a solution for selecting fast kernel
implementations in the global context by formulating it as a PBQP 
problem. Boda\cite{boda} is a code generator that generates code for CNN layers on mobile
platforms. Latte\cite{latte} is a DSL system for DNN allowing users to specify, synthesize
and optimize code for NN layers. SLINGEN\cite{slingen} is another DSL system which takes 
mathematical specifications and generates optimized C functions for linear algebra
operators with small input sizes.
These research are relevant but complementatory to our work.
\section{Conclusion and Future Work}
\label{section:conclusion}
In this paper we propose \emph{FusionStitching}, a deep fusion and code generation system
based on the XLA compilation framework for Tensorflow computations. Our system features
a critical path analysis to drive fusion decisions, a novel domain specific schedule
specification and tuning mechanism for kernels, and a shared memory optimization
technique to enable composition of large kernels. Experimental results show notable
reduction of GPU kernels, and reasonable E2E performance speedups on our benchmarks.

In practical workloads, many DNN layers only have small to medium sizes. With recent
advances on DNN kernel generation, especially on powerful hardware with mixed 
precision functionality, it would be interesting to fuse DNN layers as well
and solve a global optimization problem. 
%
%

{
\small
\bibliography{paper}
\bibliographystyle{article}
}

%
%
%

\end{document}